# Centralized Radio Resource Management for 5G small cells as LSA enabler

Óscar Carrasco[1], Federico Miatton[1], Salva Díaz[1], Uwe Herzog[2], Valerio Frascolla[3], Michael Fitch[4], Keith Briggs[4], Benoit Miscopein[5], Antonio de Domenico[5], Andreas Georgakopoulos[6]

[1]Sistelbanda SA, Valencia, Spain; [2]EURESCOM, Heidelberg, Germany; [3]Intel, Neubiberg, Germany; [4]BT, Ipswich, UK; [5]CEA Leti, Grenoble, France; [6]WINGS ICT Solutions, Athens, Greece.

*Abstract*— The stringent requirements defined for 5G systems drive the need to promote new paradigms to the existing cellular networks. Dense and ultra-dense networks based on small cells, together with new spectrum sharing schemes seem to be key enabling technologies for emerging 5G mobile networks. This article explores the vision of the SPEED-5G project, analyzing the ability of a Centralized Radio Resource Management entity to support the Licensed Shared Access spectrum sharing framework in a deployment based on Network Slicing and Network Sharing paradigms.

## I. INTRODUCTION

There is a wide consensus about the requirements that 5G systems should meet in the near future, where applications coming from diverse use cases and verticals will demand a huge increase in available bandwidth, latency reduction and a better Quality of Experience (QoE), guaranteeing a true ubiquitous broadband coverage and seamless connectivity. Besides the proposed improvements for enhancing the performance of the candidate 5G Radio Access Technologies (RATs), the research community, the industry and the regulatory bodies have identified spectrum sharing as one of the most promising ways for squeezing the potential capacity of 5G systems by means of improving the utilization of the available resources. The SPEED-5G project [1] focuses its research in the development of a three-dimensional model which allows increasing the available capacity: ultra-densification through small cells, additional spectrum and exploitation of resource across heterogeneous technologies. This three-dimensional model is referred to as extended Dynamic Spectrum Allocation (eDSA), where several spectrum bands, cells and technologies are jointly managed, so to deliver an enhanced QoE to mobile users in heterogeneous networks (HetNets). Following this model, the ability of using licensed, un-licensed and shared spectrum is the cornerstone of SPEED-5G technologies, where the optimal integration of licensed shared spectrum bands is a key aspect in order to provide future adaptability to new sharing models implemented by National Regulator Authorities.

The rest of the paper is structured as follows: section II describes the spectrum management trends and potential enhancements for current Licensed Shared Access (LSA) solutions in the framework of 5G networks, and section III introduces SPEED-5G system architecture and protocol stack, detailing SPEED-5G's centralized radio resource management (cRRM) function in relation to existing and evolved LSA architectures. Section IV provides the analysis of the cRRM solution envisioned by SPEED-5G for tackling the small cell use case for LSA, as well as introducing other important aspects like network sharing, and the advantages of using a centralized RRM entity. Finally, Section V draws the conclusions and suggests new research topics which the authors consider valuable for the commercial implementation of LSA throughout the deployment of 5G small cells in future networks.

## II. SPECTRUM MANAGEMENT AND LSA IN 5G NETWORKS

Taking into consideration the prevalent trends in spectrum management and regulation activities, shared use of licensed spectrum is one of the main topics under consideration nowadays, being actively stimulated in the different regulatory areas, including the U.S., China and Europe, where different nation-wide initiatives like the Citizens Broadband Radio Service (CBRS) in the 3.5GHz band [2] have been promoted for enabling new spectrum for broadband services. 5G requirements have been mapped into different technical challenges for future heterogeneous wireless networks in various domains [3], the following ones being the most relevant for the purpose of this analysis: the allocation of additional spectrum bands, spectral efficiency gains, network capacity enhancements, increased cost efficiency and real-time optimal resource utilization.

The 5G standardization activities consider the spectrum management aspects a central pillar of future 5G networks, but the search for new spectrum bands, both in bands above 6GHz, and in sub 6GHz bands, needs to define novel spectrum sharing models which can foster global standards adoption, enable faster access to high quality harmonized spectrum and provide a cost-efficient way to achieve economies of scale.

The envisioned 5G systems will support dynamic sharing of spectrum bands as an efficient way to overcome spectrum under-utilization, promoting the resources sharing in new spectrum bands not currently used for broadband access or future 5G services in general. The future networks need to meet the up-coming dramatic traffic growth. The 5G story and its key 5G drivers can evolve from the "use it or lose it" concept to the "use it or share it" one [4].

Within these activities, LSA appears as the European initiative for regulating the spectrum sharing framework. In that sense, LSA is defined as a complementary spectrum management tool that fits under an "individual licensing regime", a fact which allows fine management of network deployment and an effective control of the sharing arrangement, as opposed to license-exempt regulatory approaches [5]. In LSA, individual license authorizations allow a predictable management of network deployment and effective control of the sharing arrangement, having detailed information of current and future availability of the spectrum band, for the whole duration of the sharing framework. The LSA paradigm aims at ensuring a certain level of guarantee in terms of spectrum access and protection against harmful interference for both the incumbent(s) and the LSA licensees, thus allowing them to provide a predictable Quality of Service (QoS). Incumbent(s) and LSA licensee(s) obtain exclusive access to the spectrum at a given location and at a given time, being guaranteed exclusivity rights over a well-defined set of spectrum resources. Note that LSA excludes concepts such as "opportunistic spectrum access", "secondary use" or "secondary service" where the applicant receives no protection from primary users.

The current LSA system architecture, defined by the ETSI, the Radio Spectrum Policy Group (RSPG), and by the Electronic Communications Committee (ECC/CEPT), is basically addressing a static resource sharing, whereas novel LSA proposals are looking for enhancements that can lead to support more dynamic resource allocation scenarios.

This system architecture is illustrated in Fig. 1 where the main network functions and LSA players are summarized in order to provide an overall view of an end-to-end LSA system. These functional blocks are mandatory when a LSA system is implemented on a national basis [5].

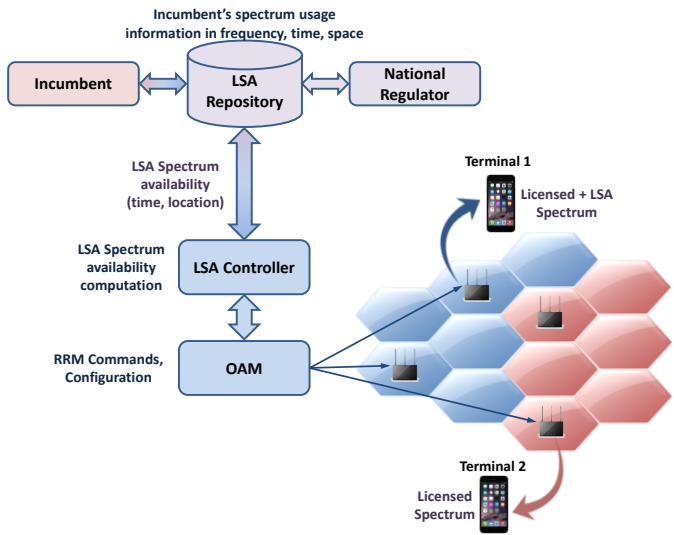

Fig. 1. *LSA Network Architecture*

The LSA repository is required to deliver information on spectrum availability and associated QoS conditions when this information is subject to changes over time. The LSA repository may be managed by an administration, the National Regulation Authority (NRA) or the incumbent, or be delegated to a trusted third party.

The LSA controller manages the access to the spectrum made available to the LSA licensee based on sharing rules and information on the incumbent's use provided by the LSA repository. The LSA controller retrieves information about spectrum from the repository through a secure and reliable communication path.

The LSA controller can interface with one or multiple LSA repositories as well as with one or multiple LSA licensee's network. There could be one or more repositories and/or controllers per country, depending e.g. on the LSA band and the incumbents' nature, but one limitation of this method is the inability of LSA repositories to share information, e.g. about spectrum usage and quality.

This is the starting point for LSA deployments, and live LSA implementations, like the LSA Finnish Trial [6][7], are basically following this functional architecture.

Nevertheless, LSA is also evolving in 5G for supporting more dynamical scenarios with different incumbent users and LSA licensees, enhancing the interaction amongst the LSA players. Additional network entities provide a new grade of coordination, guaranteeing the exclusive use of the band on a primary basis for the incumbent user (e.g. military users, radar and broadcast services), and under the LSA regime for the LSA licensees (e.g. mobile operators). Dynamic access under the LSA regime can be offered in a non-interfering basis in the LSA band, with predictable and guaranteed QoS.

The new entities proposed in [4], like the Radio Coverage Map, which has a complete knowledge of the frequencies within the LSA band that are suffering excessive interference and cannot provide the required QoS to the LSA licensees, or the Spectrum Sensing Network, which provides interference measurements from different sources, can be seamlessly integrated with the SPEED-5G architecture, setting the pace for richer LSA spectrum sharing models, as it is discussed in the following sections.

III. SPEED-5G CENTRALIZED RADIO RESOURCE MANAGEMENT

The SPEED-5G's Centralised Radio Resource Management (cRRM) is a key element of SPEED-5G's 5G small cell network architecture and Protocol Stack. The cRRM has the task of applying the spectrum management policies defined by the regulation authorities, selecting the most appropriate spectrum band and RAT to be used by each session.

Fig. 2 shows the SPEED-5G's Protocol Stack, depicting a centralised RRM entity deployed in an edge cloud, having a cluster of different 5G small cells under its control.

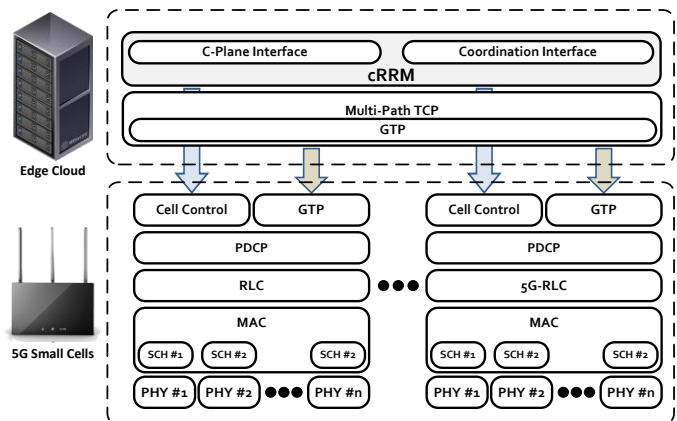

Fig. 2. *SPEED-5G Protocol Stack*

In this edge cloud, the cRRM entity manages multi-user, multi-cell and multi-connection network capacity issues. It implements operator's strategies and algorithms for mobility and link management, such as dynamic spectrum access, carrier aggregation, interference coordination, inter-RAT load balancing, and handovers. The cRRM also processes the UE measurement reports in order to trigger the signaling of the procedures related with the air interface, including the configuration of the different radio bearers, and other procedures related to QoS management. Indeed, this entity is mainly responsible for ensuring the provision of the end-to-end QoE, by aggregating multiple bearers flowing across multiple heterogeneous RATs, possibly operating in a very diverse range of frequency bands having different characteristics.

Therefore the cRRM entity is mainly responsible for maximizing the system spectral efficiency across the different available frequency bands. Additionally, multiple instances of cRRM are interconnected with one another in order to coordinate different clusters of small cells at the edge of their respective geographical areas of coverage.

This function also includes the intelligence for configuring the different layers of the Protocol Stack of the air interface of the 5G Small Cell network, cooperating with the LSA Controller and properly applying LSA sharing policies and transmission grants at the LSA spectrum.

The protocol stack of Fig. 2 also includes a Multipath TCP layer, an essential component for scheduling different TCP services over different radio bearers in parallel. These radio bearers are embedded in different data tunnels, flowing across one or more RATs of one or multiple cells. In combination with the traffic steering capabilities of the cRRM layer, the Multipath TCP layer is the main enabler to support the scheduling of TCP services over multiple carriers in parallel enabling multi-cell coordination schemes like Cooperative Multi-Point (CoMP) and virtual multi-cell MIMO, configuring an edge cluster as a distributed massive MIMO supercell.

This protocol stack proposal is fully aligned with novel 5G architecture proposals, where the processing of signaling data is moved to an edge cloud, and the service anchor and content delivery are localized, achieving local routing and backhaul capacity savings [8][9]. Additionally, this protocol stack implements the SPEED-5G enhancements related with dynamic spectrum access and LSA, that constitute two key enablers for reaching the capacity gains envisioned by next-generation 5G systems.

The benefits of having a centralized RRM entity are manifold. In the following, the main advantages of having a cRRM as key enabler for LSA are briefly reviewed.

*A. Single point of interaction with the LSA Repository*

The cRRM is the single point of interaction with the LSA Repository via the LSA Controller. Typically, a small cell network consists of multiple heterogeneous nodes, and the management of purely distributed solutions relies on interacting with an Element Management System using a TR-069 management interface. Such approach needs a clear definition of standardized interfaces across heterogeneous systems, and brings issues related to the integration of solutions from multiple vendors. Having a single centralized entity simplifies the integration and the management of multiple heterogeneous systems.

*B. Better spectrum exploitation across the space dimension*

Another advantage brought by this network architecture is the possibility to diversify and optimize the spectrum usage in the space dimension. The advantage of having a centralized cRRM permits embedding the space dimension into the decisions related to spectrum allocation. The availability of certain frequency bands can be highly location-dependent, as is the case for the 2.6GHz frequency band used by maritime radars which operate only in coastal regions. This use-case can be easily supported by the SPEED-5G architecture previously introduced, as this is a typical information that is available to the LSA Repository (and hence to the cRRM) but not directly to the small cells.

*C. Cell coordination at the TTI level*

The smallest time granularity at which the cRRM may operate can be as fine as one single Time Transmission Interval (TTI). The operations that can be performed at such resolution include resource allocation and reconfiguration, as well as the application of any other network optimization techniques. The ability to perform cell coordination decisions per TTI leads to a better interference management and higher QoE gains.

*D. Time synchronization accuracy to perform resource sharing across multiple operators*

Radio sharing is a new concept recently introduced to achieve optimal spectrum usage by sharing on demand the limited radio resources by means of the small cells. A necessary condition to enable radio resource sharing across multiple network operators is the ability to provide accurate time synchronization between the different core networks. In fact, this is fundamental in order to meet potential Service Level Agreements (SLA) signed with the different network operators, which act as LSA licensees.

*E. Centralized interference management*

The cRRM operates as a centralized point of management to control a cluster of small cells. Having such a centralized entity permits simplification of interference management operations, paving the way for the implementation of efficient algorithms to perform load balancing across heterogeneous bands and multiple RATs.

*F. Reduction of latency requirements due to the proximity of deployment of the cRRM to the small cell*

The deployment of the cRRM in an edge gateway operating as a single decision point for a cluster of small cells reduces the delays associated with the process of decisions making compared to the case when the RRM is implemented remotely in the cloud. This allows the deployment and implementation of techniques that operate on relatively small time-scales, which is particularly critical for mechanisms such as interference management and dynamic spectrum allocation that must respond dynamically to rapid changes in the surrounding environment.

The cRRM emerges as the most important element to support the ultra-densification of small cells that is envisioned for 5G networks. Although ultra-dense networks could in principle reduce the spectrum efficiency due to the reduction of Signal-to-Interference-plus-Noise Ratio (SINR) caused by higher interference levels, the introduction of novel techniques for interference management and for a better exploitation of the spectrum will counteract these negative effects.

Note that these clever techniques will all boil down to creating an intelligent network that is able to respond fast to the dynamic variations in the environment. Example of relevant works in this area include a whole area of research devoted to small cells enabled with cognitive capabilities such as [10][11][12] on the one hand, and related to advanced dynamic channel selection techniques on the other hand, such as [13][14][15][16]. The key assumption for all these cases is that the small cells are equipped with sensing capabilities that enable them to measure the interference levels and to evaluate the spectrum usage across the different available bands.

In the present section, the characteristics of having a centralized RRM were provided, together with the main advantages of the cRRM as the key enabler to support the LSA paradigm. The SPEED-5G project intends to go beyond the current state-of-the-art by envisioning intelligent small cells able to support multiple RATs and heterogeneous bands [17]. The next section describes the 5G small cells envisioned by the SPEED-5G project.

## IV. LSA SMALL CELL USE CASE

As it has been identified by different trials and LSA research projects [4][7], a small cell deployment is one of the most significant LSA use cases to be analyzed, as capacity improvements in dense populated scenarios, which can be found in U.S., Asian and European cities.

In LTE-A HetNet deployments, small cells can be found both indoors and outdoors, with and without macro-cell overlay coverage, with both uniform and non-uniform traffic load distribution in time-domain and spatial-domains, with and without ideal backhaul [18]. In this HetNet scenario, 3GPP proposes the adoption of the 3.5GHz band, an LSA band candidate, as an additional small cell band which would allow the use of wider bandwidths (up to 100MHz), the support of dual cell connectivity and 256-QAM modulation short-range radio links. For aligning this scenario with 5G use cases, defining thus the 5G LSA small cell scenario, we include an additional requirement, indoor small cells multi-tenancy through the Multi-Operator Core Network Radio Access Network (MOCN RAN) sharing scheme, in order to minimize the investment cost for deploying small cells in indoors, following end user's rationale which is in line with having one cell connected to multiple operators avoiding deploying one small cell per operator.

The proposed evolved LSA network architecture with the different LSA players and needed network functions, including both the Edge Cloud with the cRRM and the LSA functions, is depicted in Fig.3.

The use case is defined as follows: both operators are LSA licensees, having exclusivity rights over an LSA sub-band, but they have agreed on a voluntary basis to share the network infrastructure through a MOCN RAN sharing scheme and to outsource the Edge Cloud management to a Neutral Host operator, which implements the sharing policy related with LSA spectrum utilization, ensuring that both operators will have, at least, the same QoE over the LSA band than they can get individually due to the statistical gain of the concurrent access.

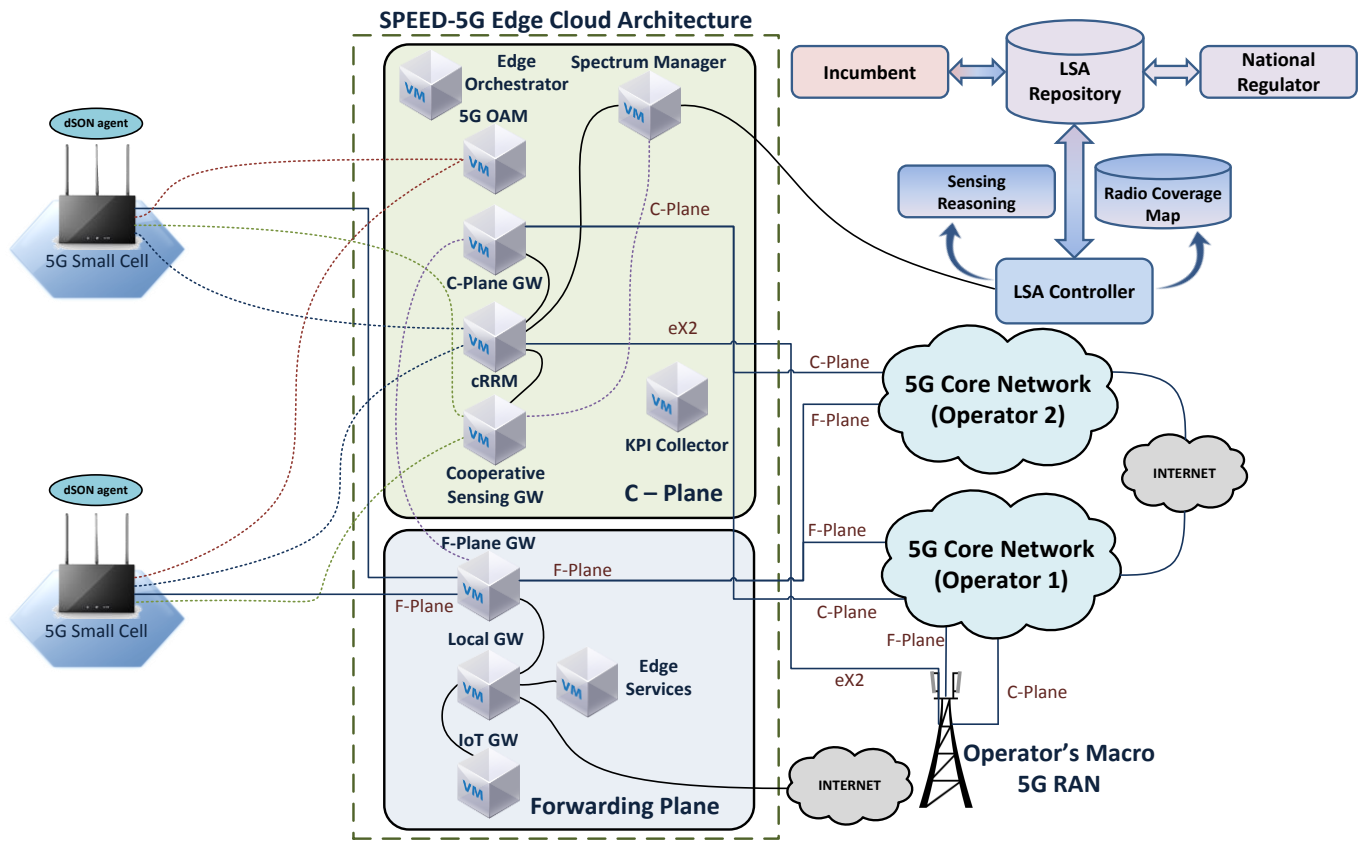

*Fig.3 SPEED-5G's Edge Cloud architecture integrated with an evolved LSA system for the LSA small cell Use Case*

In this network implementation, the cRRM is the key element for enabling a rich and dynamic LSA framework, allowing optimizing spectrum allocation and overall QoE, as it is discussed in the following paragraphs.

Real-time management interface: typical indoor small cell networks are managed using a TR-069 interface, a heavy Layer-3 oriented interface which is not intended for supporting real-time operation, having typical reporting periods between 15-60 minutes, being not suitable for providing short-term reports needed for key aspects like interference management. The cRRM – Cell Control interface is a Layer-2 oriented interface, using Abstract Syntax Notation One (ASN.1) or other efficient coding scheme that allows having very short latencies in comparison to xml-based interfaces.

The cRRM entity enables a Cloud-RAN like operation by means of centralizing the network intelligence and decision-making for self-optimization and control plane management. The cloud edge allows efficiently managing ultra-dense network resources defining an edge cloud as a pool of virtual network resources, allowing as well new business and ownership models without the typical stringent latency requirements of Cloud-RAN deployments.

Another advantage of having a cRRM as a single management entity for a cluster of small cells is related to predictable QoS indicators that operators are able to manage. This possibility is achieved on one hand through the centralization of the spectrum resources allocation performed at the cRRM, and on the other hand by means of advanced interference coordination mechanisms, such as Inter-cell interference coordination (ICIC), enhanced Inter-cell interference coordination (eICIC), and Dynamic Frequency Selection (DFS).

Furthermore, the envisioned ultra-dense network of small cells makes it possible to deploy a cooperative and coordinated sensing network where the small cells themselves act as the main sensing entities, implementing a Cognitive Wireless Cloud [10][11][12]. This in turn enables the creation of an accurate, up-to-date and almost real-time SINR maps, which is crucial information for achieving the optimal usage of the LSA band. Note that this is possible only thanks to the defined architecture based on cRRM as a central management entity for clusters of small cells.

V. CONCLUSIONS

This paper describes how a cRRM entity together with an Edge Cloud can enable and drive the implementation of evolved LSA systems utilization of small cells. LSA and SPEED-5G cRRM solutions are explored. The analysis of the LSA small cell use case is provided identifying the aspects where a cRRM is enabling an efficient LSA framework, analysing as well the interaction of SPEED-5G architecture and an evolved LSA system architecture. Future research in LSA coordination procedures, network interfaces and sensing strategies is needed for setting a complete framework of LSA in the 5G standardization path.

ACKNOWLEDGMENT

The research leading to these results received funding from the European Commission H2020 programme under grant agreement n°671705 (SPEED-5G project).